\documentstyle[tighten,preprint,aps]{revtex}

\begin{document}
\draft

\def\no{\nonumber}
\def\a{\alpha}
\def\be{\begin{equation}}
\def\ee{\end{equation}}
\def\p{\partial}
\def\T{\Theta}
\def\r{\rho}
\def\k{\kappa}
\def\a{\alpha}
\def\b{\beta}
\def\e{\epsilon}
\def\g{\gamma}
\def\t{\theta}
\def\S{\Sigma}
\def\l{\lambda}
\def\L{\Lambda}
\def\n{\noindent}
\def\o{\omega}

\preprint{\vbox{\baselineskip=12pt
\rightline{IUCAA-40/99}
\rightline{hep-th/9911069}
}}
\title{New classes of black hole spacetimes in 2+1 gravity} 
\author{Sukanta Bose \footnote{Electronic address: 
{\em Sukanta.Bose@astro.cf.ac.uk}}
${}^{(1)}$, Naresh Dadhich\footnote{Electronic address: {\em nkd@iucaa.ernet.in
}}${}^{(1)}$, and Sayan Kar\footnote{Electronic address: {\em 
sayan@phy.iitkgp.ernet.in}}${}^{(2)}$}
\address{{\rm (1)}Inter-University Centre for Astronomy and Astrophysics, Post 
Bag 4,Ganeshkhind,\\ Pune 411007, India}
\address{{\rm (2)}Department of Physics and Centre for
Theoretical Studies \\Indian Institute of Technology, Kharagpur 721 302, India}
\date{September 1999}
\maketitle


\begin{abstract}
The electrogravity transformation is applied to the three-dimensional 
Einstein field equations to obtain new multi-parameter families of black hole 
solutions. The Ba\~{n}ados-Teitelboim-Zanelli black hole is shown to be a 
special case of one of these families. The causal structure, associated matter,
as well as the mechanical and thermodynamical properties of some of the 
solutions are discussed. 

\end{abstract}

\pacs{04.70.Dy, 04.62.+v,11.10.Kk}


Three-dimensional (3D) gravity provides an excellent arena for learning about 
quantum effects of gravity in a simplified context \cite{3dqg}. 
It helps that this theory, with a negative cosmological constant, 
$\Lambda$, has non-trivial solutions, such as the 
Ba\~{n}ados-Teitelboim-Zanelli (BTZ) black hole spacetime\cite{BTZ}, which 
provide important testing grounds for such studies.
A few other 3D black hole solutions have also been found by coupling matter 
fields to gravity in different ways \cite{2dbh}. 

However, non-trivial solutions in 3D are relatively less compared to general
relativity (GR) \cite{DJ}. This is due to the lack of gravitational 
interaction in the theory: unlike its 4D counterpart, localized matter here 
cannot by itself give rise to many interesting solutions, such as black holes.
In this paper we report the finding of two multi-parameter families of black
hole solutions in 3D Einstein gravity, with (and without) a negative $\Lambda$,
and in the presence of different types of matter. One of these families 
includes the BTZ hole as a particular case. We use the ``electrogravity'' 
transformation (EGT), recently proposed in the context of GR 
\cite{ND1,ND2,DP1,NDL}, to obtain our solutions.

The EGT is defined as the interchange of 
active and passive electric parts of the Riemann curvature tensor. This 
exactly translates into the interchange of Ricci and Einstein tensors 
{\cite{ND1}}. It is a duality transformation for the vacuum field equations 
since it leaves them invariant. In this work, a pair of spacetimes will be 
termed as related by EGT if they arise as solutions to the 
equations $F_i (G_{ab})=0$ and $F_i (R_{ab}) = 0$, respectively, where $F_i$ 
denotes a pre-defined set of functions. Note that these spacetimes may arise 
as solutions to different matter distributions, in general. However, their 
choice can be restricted by suitably specifying the forms of $F_i$ and, 
therefore, the field equations that they correspond to (as discussed below).
By a common abuse of nomenclature, equations related by EGT and, therefore, 
their respective solutions, will be termed as ``dual'' to each other. 
We follow the conventions of Wald \cite{RMW} and employ the units, 
$G=1/8$ and $c=1$. Specifically, the metric signature for 3D spacetimes is 
$(-,+,+)$. The convention for the totally antisymmetric 
Levi-Civita tensor, $\eta_{abc}$, is such that, $\eta_{012} = 1 $.

Following {\cite{ND1}}, we define the active electric part of the Riemann 
tensor relative to a timelike unit vector as, $E_{ac} = R_{abcd} u^b u^d$, 
which is a purely spatial, symmetric tensor. The double-dual of the Riemann 
tensor in 3D, namely, $*R*_{ac}$, is defined as
$*R*_a {}^c =  {1\over 4}{\eta}_{amn} \eta^{cpq} R^{mn}{}_{pq}$,
which is just the Einstein tensor $G_a{}^c$. It can be used to construct the 
following three-vector:
$G_a = G_a{}^c u_c = {1\over 4}{\eta}_{amn} \eta^{cpq} R^{mn}{}_{pq} u_c $.
The spatial projection of $G_a$ is the 3D analogue of the magnetic 
part. It is defined as ${\tilde{E}_a} =  G_b h^b{}_a $,
where $h_{ab}$ is the spatial metric, $h_{ab} = g_{ab} + u_a u_b $, 
with $g_{ab}$ being the spacetime metric and $h_{ab} u^a =0$. Thus, $E_{ac}$ 
and $\tilde{E}_a$ account for five of the 
six independent components of the Riemann tensor. The remaining component is 
the temporal part of $G_a$, namely, $G_a u^a$, which is the 3D analogue of
the passive electric part.

The Ricci tensor can be decomposed in terms of $E_{ab}$, $G_{ab}$,
and their various projections as follows:
\be\label{Rdec}
R_{ab} = -E_{ab} + (G_c u^c) h_{ab} + Eu_a u_b - (\tilde{E}_a u_b 
+ \tilde{E}_b u_a) \ \ ,
\ee
where $E$ is the trace of $E_{ab}$. On the other hand, the Einstein tensor can
be written as 
\be \label{Gab}
G_{ab} = -E_{ab}+Eh_{ab} +(G_{c}u^c)u_au_b - (\tilde{E}_a u_{b}+
{\tilde{E}}_{b}u_a) \,.
\ee
The transformation
\be\label{dt1}
E \longleftrightarrow (G_c u^c) \ \ ,
\ee
interchanges the Ricci and the Einstein tensors and, therefore, effects an
EGT. Requiring that the spatial and temporal components of the vacuum field 
equations, $R_{ab} = 0$, vanish separately, gives:
\begin{mathletters}%
\label{EBeom}
\begin{eqnarray}
-E_{ab} +(G_c u^c) h_{ab} &=& 0 \ \ ,\label{EBeom3} \\
\tilde{E}_a &=& 0 \ \ , \quad {\rm and}  \label{EBeom2} \\
{\rm either}\quad E = 0 \quad &{\rm or}& \quad (G_c u^c) =0 \label{EBeom1} \,.
\end{eqnarray}
\end{mathletters}%
The above set of six equations is invariant under EGT. Flat spacetime is the 
unique solution of these equations. 

Consider a circularly symmetric spacetime (CSS) with the line element:
\begin{equation}
ds^{2} = -a(r)dt^{2} + b(r)dr^{2} + c(r)d\phi^{2} \,.
\end{equation}
If the field equations,
\be \label{Rvac}
{R_2}^2 =0\>, \quad {R_1}^1= {R_0}^0 \ \ ,
\ee
are obeyed everywhere, then it implies that such a spacetime with $c=r^2$ 
(or, alternatively, $ab=1$) has to be flat (vacuum). The above field equations
are less restrictive in form than the equation $R_{ab} =0$. In fact they 
follow from a simple requirement on the matter energy density, $\rho = T_{ab}
u^a u^b$, and the ``convergence density'' of a family of radial 
null-geodesics, $\rho_n \equiv T_{ab} k^a k^b$, with $k^a$ being the tangent
along such a geodesic; it is:
\be\label{rhocond}
\rho =0 \>, \quad {\rm and} \quad \rho_n =0 \ \ ,
\ee
everywhere. Also the convergence density of a family of timelike geodesics is
defined as $\rho_t = (T_{ab} -g_{ab} T/2) u^a u^b$.

Since the EGT implies $R_{ab} \longleftrightarrow  {G_{ab}}$,
its action on Eq. (\ref{Rvac}) gives,
\be \label{Gvac}
G_2{}^2=0 \>, \quad {G_1}^1 = G_0{}^0 \,.
\ee
Alternatively, the above equations can be obtained by implementing the
transformation $\rho \to \rho_t$, $\rho_n \to \rho_n$ on the equations
(\ref{rhocond}). For $c=r^2$ (or, alternatively, $ab=1$), the above equations
imply: 
\be \label{Gvacsol}
a= \a r +\b  = b^{-1} \ \ ,
\ee
for $r >0$. Here $\a$ and $\b$ are integration constants. For $\a =0$ and 
$\b >0$, the above solution is a locally flat spacetime with a deficit angle. 
It describes the spacetime region around a massive point particle \cite{DJ}. 
It can also describe a region of spacetime outside an electromagnetic wave 
with compact support, which arises as a solution to the 3D Einstein-Maxwell 
system \cite{AA}. On the other hand, if $\b<0$, then in the limit $\a\to 0$, 
the resulting spacetime is identical to the spinless BTZ for very large radius 
of curvature (i.e., $\Lambda\to 0$). Such a spacetime describes an expanding 
cylinder. As we now show, Eq. (\ref{Gvacsol}) represents a 
non-vacuum spacetime.

The corresponding matter distribution is given by the stress tensor 
components:
\be\label{dmRT01}
T_0{}^0 = T_1{}^1 = \a/ (2r) + \pi \b\delta (r)/r \ \ ,
\ee
with all other components vanishing. The first term arises from string-dust 
matter (see below), whereas the second term indicates a point source at the 
center of symmetry.\footnote{This is an interesting deviation from 4D gravity:
it is well known that the Schwarzschild spacetime can not arise as a solution 
to a point source, unlike here in 3D. This difference is again due to the 
absence of gravitational interaction in 3D.}

In the presence of $\Lambda$, the field equations are:
\be\label{EFL}
R_{ab} = {2\L}g_{ab} \equiv -{2\over l^2}g_{ab} \ \ , 
\ee
where $\Lambda = -l^{-2}<0$. Thus,
Eqs. (\ref{Rvac}) get modified to:
\be \label{RL}
{R_2}^2 =-2/ l^2\>, \quad {R_1}^1= {R_0}^0 \,.
\ee
With $c=r^2$ (or, alternatively, $ab=1$), these equations yield the spinless 
BTZ spacetime  \cite{BTZ} as the unique CSS solution.
Its metric is given by
\be\label{BTZmet}
ds^2 =- a_{\rm BTZ} dt^2 +a_{\rm BTZ}^{-1} dr^2 +r^2 d\phi^2 \ \ ,
\ee
where $a_{\rm BTZ}=r^2 /l^2 -M$ and $M$, an integration constant, is a 
conserved charge associated with 
asymptotic invariance under temporal displacements. When $M>0$, it can be 
interpreted as the black hole mass parameter. In such a case, the above 
spacetime is asymptotically anti-de Sitter (AdS). Setting $M=-1$ gives the 
AdS spacetime itself. The above metric can also be generalized to include 
rotation \cite{BTZ}.

Under EGT, Eqs. (\ref{RL}) go over to
\be\label{GL}
{G_2}^2= -2/ l^2 \>,\quad {G_1}^1 = {G_0}^0 \,.
\ee
The above set is solved (for $c=r^{2}$, $r > 0$) by:  
\be\label{dmRLsol}
a = -2r^2/ l^2 +\a r +\b =b^{-1} \,.
\ee
Similarly, if $l^2 <0$ in Eqs. (\ref{RL}), then the solution to its dual set 
(Eqs. (\ref{GL}) with $l^2<0$) has the line element:
\be\label{dmRLsolmet}
ds^2 = -\bar{a} dt^2 + \bar{a}^{-1}dr^2 +r^2 d\phi^2 \ \ ,
\ee
where $\bar{a} =r^2/\bar{l}^2 +\a r +\b$. We have 
defined ${\bar l}^{2}=-l^2 /2 $. The matter distribution 
producing the above metric is the same as that given by Eqs. (\ref{dmRT01}).
In the above metric one can include rotation to obtain a four-parameter
family of black hole solutions:
\begin{equation}\label{metJ0}
ds^2 =-a_{\rm rot} (r) dt^2 +a_{\rm rot}^{-1} (r) dr^2 +
r^2 \left ( d\phi -\frac{J}{2r^{2}}dt \right )^{2} \ \ ,
\end{equation}
where $a_{\rm rot}=r^2/ \bar{l}^2 +\a r-M +J^2/(4r^2 )$ and $J$ is the 
angular momentum. Also, we set $\b=-M$.
We now explicitly show that the members of the subfamily with 
$J=0$ have the causal structure of a black hole. 

The maximal extension of the above spacetime (with $J=0$) can be achieved by 
transforming to the analogue of the Kruskal double-null coordinates,
\begin{eqnarray}
U&=&-{1\over \a \bar{l}} e^{-\a\bar{l}u} \equiv -{1\over \a \bar{l}}\exp\left[ 
-\a\bar{l} \left(t-r_*\right)\right]\quad{\rm and} \nonumber\\
V&=&{1\over \a \bar{l}} e^{\a\bar{l}v} \equiv {1\over \a \bar{l}}\exp\left[\a
\bar{l} \left(t+r_*\right)\right]  \ \ , \label{Kcoord}
\end{eqnarray}
where $r_*$ is the equivalent of the tortoise coordinate:
\be \label{tortr}
r_* = {1\over \bar{l}\sqrt{\a^2\bar{l}^2+4M}}
\ln \left[ {r-r_H\over r+r_H+\a\bar{l}^2} \right] \,. 
\ee
Above, $r = r_H$ is the location of the horizon. It is related to the
parameters of the solution by,
\be\label{rH}
r_H = \bar{l}^2 (-\a+\sqrt{\a^2+4M/\bar{l}^2})/2 \,.
\ee
In terms of the Kruskal double-null coordinates, the metric takes the 
following form:
\be\label{metJ0K}
ds^2 = \left({r\over\bar{l}} +{\a\bar{l}\over 2} +\sqrt{{\a^2\bar{l}^2\over 4}
+M} \right) dUdV + r^2 d\phi^2 \ \ ,
\ee
which is clearly non-singular at the horizon. The singularity itself lies at 
$r=0$. This can be inferred from the behavior of the Ricci scalar, which is
\be
R = -{6/ \bar{l}^2} -{2\a/ r} \,.
\ee
Note that for $\a =0$, the metric is locally AdS. However, for $M\geq 0$, it 
describes a BTZ black hole. For $J=0$ and $1/\bar{l}=0$, the metric 
(\ref{metJ0}) is the same as the dual-vacuum solution (\ref{Gvacsol}), with 
$\beta$ replaced by $-M$. The causal structure of (\ref{Gvacsol}) can be 
easily recovered following a similar analysis as above. For $\alpha >0$ and 
$M >0$, it is a black hole with a horizon at $r= M/\a$ that cloaks a spacelike 
singularity at $r=0$. 

The matter that  yields the line element 
(\ref{metJ0}) corresponds to a ``cloud'' of strings
\cite{PSL}, apart from the point source at the center \cite{footnote}. 
The associated stress tensor is:
\be\label{TSsd}
T_{\rm string}^{\mu\nu} = \varrho \Sigma^{\mu\beta} \Sigma^\nu_\b / (-\sigma)^{
1/2} \ \ , \ee
where $\varrho$ is the proper energy density of the cloud, $\sigma_{\mu\nu}$ 
is the 2D metric on a string world-sheet, and $\Sigma^{\mu
\nu}$ is the bivector associated with this world-sheet:
$\Sigma^{\mu\nu} = \epsilon^{AB} {\partial x^\mu\over \partial \l^A}
{\partial x^\nu\over \partial \l^B} \,.
$
Here, $\epsilon^{AB}$ is the 2D Levi-Civita tensor (normalized such that
$\epsilon^{01}=-\epsilon^{10}=1$) and $\l^A = (\l^0, \l^1)$, where $\l^0$ and
$\l^1$ are a timelike and a spacelike parameter on the string world-sheet.
Following Ref. \cite{PSL} for 3D static CSS, it can be showen
that the only non-zero components of the string-dust stress 
tensor (\ref{TSsd}) are,
\be\label{TSsdcomps}
T_{{\rm string}\>0}{}^0 = T_{{\rm string}\>1}{}^1 \propto r^{-1}  
\ \ ,\ee
which are associated with the dual solutions.

The matter stress tensor (\ref{dmRT01}) corresponding to the dual-vacuum
solution (\ref{Gvacsol}) contains a couple of parameters, $\a$ and $\b$. 
Matter with $\a>0$ and $\b<0$ is associated with a black hole.
By itself, the string-dust part of this matter stress tensor violates the null 
energy condition (NEC), whereas the point-source term obeys it
at $r=0$. For $\a<0$ and $\b> 0$, one finds a naked singularity at 
$r=0$ and a horizon similar to a cosmological horizon in de Sitter space. 
However, even though the string-dust part of the matter stress tensor obeys 
NEC here, the point-source term violates it, in conformity with cosmic 
censorship. For both $\a$ and $\b$ positive, the spacetime contains a naked 
singularity, but no horizon. Also, the NEC is violated everywhere in this case.

The non-static CSS (\ref{metJ0}) can also be 
produced with a string-dust distribution. Here, however, not only 
$\Sigma^{01}$, but also $\Sigma^{12}$ is non-zero. The corresponding stress 
tensor can be cast in the form given in Eq. (\ref{TSsdcomps}), with the 
indices now referring to the orthonormal triads rather than the curvature 
coordinates.

We now discuss some mechanical and thermodynamical aspects of the 
solution in Eq. (\ref{metJ0}). In keeping with BTZ, we choose the ground
state of this solution to have the same metric, but with $M=0$ and $J=0$. 
Then, $M$ and $J$ are just the Noether charges associated with asymptotic
invariance under Killing time-displacements and rotations. They are the mass 
and angular momentum of the hole, respectively. Indeed, they can be identified 
with the Brown and York (BY) quasilocal mass and angular momentum as well, 
when the quasilocal surface is taken at infinity \cite{BY,BCM,BL} and the 
reference spacetime is taken to be the ground state.

The entropy of a stationary black hole in Einstein gravity depends only on the
horizon geometry. In 3D it is twice the horizon circumference independent of
the asymptotic behavior of gravitational fields and the presence of matter 
fields\cite{BTZ,BCM}. Thus, the entropy of our solution (\ref{metJ0}), with 
$J=0$, is $4\pi r_H$. Also, for such a hole the surface $r=r_H$ is a Killing 
horizon. The Killing vector normal to this surface is $\chi = \partial_v$. 
Thus the surface gravity, $\kappa$, is given by 
$\kappa^2 = -\nabla^a\chi^b\nabla_a\chi_b /2$.
At the horizon, $\kappa_H =\sqrt{\a^2 +4M/\bar{l}^2} /2$. (Interestingly, for
metric (\ref{Gvacsol}) $\kappa=\a/2$ everywhere and, thus, is an 
$M$-independent constant in 
space and time.) For a canonical ensemble of such black holes, the 
thermodynamic internal energy can be identified with the BY quasilocal energy 
inside a circular ``box'' of curvature radius $r=R$:
\be
E= -2\left[\sqrt{R^2/\bar{l}^2 +\a R -M} -\sqrt{R^2/\bar{l}^2 +\a R}\right] \,.
\ee
Then the temperature on the box is
\be
T(R)=\left({\partial E \over \partial S}\right)= {1\over 4\pi}\sqrt{\a^2 +
4M/\bar{l}^2 \over R^2/\bar{l}^2 +\a R-M} = {\kappa_H \over2\pi N(R)} \ \ ,
\ee
where the lapse $N(R)$ causes the temperature to redshift with distance. The 
last expression has exactly the same form as that for a BTZ black hole 
\cite{BCM}.

We now consider the case of ``self-duality'' for EGT, i.e., $R_{ab} 
\longleftrightarrow G_{ab}$. The self-dual 
solutions obey $R_{ab}=G_{ab}$, which implies $R=0$. However, 
$R_{ab}$ need not be zero everywhere, otherwise it would imply a vacuum
solution. A particular class of self-dual solutions, which obey
$\rho_n = R_0{}^0 - R_1{}^1 = 0$, is:
\begin{equation} \label{Sch3D}
ds^{2} = -a_{\rm sd} dt^{2} +a_{\rm sd}^{-1} dr^{2} + r^{2}d\phi^{2} \ \ ,
\end{equation}
where $a_{\rm sd} = 1-C_{1}+C_{2} /r$. This describes an asymptotically flat 
spacetime with a deficit angle when $C_{1}\neq 0$. For $C_1 =0$, we call this 
solution the 3D analogue of the Schwarzschild, largely due to
its asymptotic flatness and due to the fact that for $C_2 < 0$,
it is the equatorial section of the Schwarzschild solution. We are 
not aware of any solution in 3D with these features. The 
causal structure of this metric is analogous to that of Schwarzschild. 
The $C_1 \neq 0$ metric itself describes the equatorial section of the 
Barriola-Vilenkin global monopole solution \cite{BV}.

The matter stress tensor for the self-dual solution has the form 
(for $C_{1}=0$) :
\begin{equation}
T_{0}{}^{0} = -\frac{C_{2}}{2r^{3}} \hspace{.2in} ; \hspace{.2in}
T_{1}{}^{1} = -\frac{C_{2}}{2r^{3}} \hspace{.2in} ; \hspace{.2in}
T_{2}{}^{2} = \frac{C_{2}}{r^{3}} \,.
\end{equation}
The salient feature of this stress tensor is its tracelessness. It can be
shown that massless, non-minimally coupled scalar fields are a source with 
such a stress tensor \cite{BD}. On demanding that 
it obeys NEC, one gets a naked singularity. A violation of 
NEC allows an eternal black hole. However, regardless of whether $C_{2}$ 
is negative or positive, the averaged null energy condition (ANEC) along 
radial null geodesics is always satisfied. Since solution (\ref{Sch3D}) is 
associated with a fundamental field, it presents a promising arena for studying
quantum properties of gravity coupled to matter.

The above result on self-dual solutions can be easily extended to $D>3$.
For example, in $D=4$, on solving $R=0$ (together with 
$\rho_n = 0$), one obtains the Reissner-Nordstrom solution (which of course 
includes the Schwarzschild solution as a special case). This solution is again
the equatorial section of the vacuum solution in one higher dimension, namely,
5D. This appears to be a general property of spherically symmetric solutions 
to the equation $R=0$ for $D \geq 3$.

Our results demonstrate new classes of black
hole spacetimes in 3D. They are related via EGT
to known solutions. In fact, even without involving $\Lambda$
or a non-trivial coupling dependent on, say, a dilaton or a Brans-Dicke scalar
field, it is remarkable that 3D gravity possesses a black hole, namely, the 
{\em dual} to flat spacetime, Eq. (\ref{Gvacsol}). This solution is 
associated with string-dust matter. This is true about some dual solutions in 
4D as well (see, e.g., Ref. \cite{NDL}). Thus, the association of string-dust 
with the metrics generated via EGT appears to be quite 
generic, even across spacetime dimensions. These distributions in 3D are, 
however, distinguishable from those in 4D by their different (radial) 
fall-off behavior. 
 
Finally, it is possible to generalize our asymptotically non-flat 3D
solutions further by the inclusion of circularly symmetric electromagnetic 
fields and/or other matter. Also, issues related to black hole thermodynamics 
as well as quantum gravity in the context of these new geometries remain 
interesting problems to be explored.

We would like to thank Steven Carlip for his useful comments. SK thanks IUCAA 
for hospitality.



\begin{references}
\bibitem{footnote}
Note, however, 
that in the case where $\a=0$, the resulting metric is pure BTZ and one does 
not need a point source. The ``$M$-term'' is produced by an identification of 
points in AdS through a discrete subgroup of its symmetry group $SO(2,2)$
\cite{BHTZ}.
\end{references}
\end{document}